\documentclass[preprint,a4paper]{elsarticle}

\usepackage{lineno,hyperref}
\usepackage{amsmath,amssymb,amsthm}
\usepackage[utf8]{inputenc}
\usepackage{listings}
\usepackage{xcolor}
\usepackage{pgfplots}
\usepackage{pgfplotstable}

\usepackage{stmaryrd}
		\SetSymbolFont{stmry}{bold}{U}{stmry}{m}{n}
\usepackage{placeins}

\renewcommand{\div}{\operatorname{div}}

\newcommand{\N}{\mathbb{N}}

\newcommand{\R}{\mathbb{R}}
\renewcommand{\S}{\mathbb{S}}

\newcommand{\bfc}{\mathbf{c}}

\newcommand{\bfg}{\mathbf{g}}

\newcommand{\bfj}{\mathbf{j}}

\newcommand{\bfn}{\mathbf{n}}

\newcommand{\bfp}{\mathbf{p}}

\newcommand{\bfx}{\mathbf{x}}
\newcommand{\bfy}{\mathbf{y}}

\newcommand{\bfE}{\mathbf{E}}

\newcommand{\bfH}{\mathbf{H}}

\newcommand{\bfNull}{\mathbf{0}}

\newcommand{\cH}{\mathcal{H}}

\newcommand{\cS}{\mathcal{S}}

\newcommand{\bfcS}{\boldsymbol{\mathcal{S}}}

\renewcommand{\div}{\operatorname{div}}

\DeclareMathOperator{\bcurl}{{\mathbf{curl}}}
\DeclareMathOperator{\bgrad}{{\boldsymbol{\nabla}}}
\renewcommand{\d}{\operatorname{d}\!}

\newcommand{\isdef}{\mathrel{\mathrel{\mathop:}=}}

\definecolor{mygreen}{rgb}{0,0.6,0}
\definecolor{mygray}{rgb}{0.5,0.5,0.5}
\definecolor{mymauve}{rgb}{0.58,0,0.82}

\theoremstyle{plain}             
\newtheorem{theorem}{Theorem}[section]

\newtheorem{remark}[theorem]{Remark}

\lstdefinestyle{customc}{
  belowcaptionskip=1\baselineskip,
  breaklines=true,
  frame=L,
  xleftmargin=\parindent,
  language=C++,
  showstringspaces=false,
  basicstyle=\footnotesize\ttfamily,
  keywordstyle=\bfseries\color{green!40!black},
  commentstyle=\itshape\color{purple!40!black},
  identifierstyle=\color{blue},
  stringstyle=\color{orange},
}

\journal{arXiv}

\begin{document}

\begin{frontmatter}

\title{Bembel: The Fast Isogeometric Boundary Element C++ Library for Laplace, Helmholtz, and Electric Wave Equation}

\author[aff1]{J.~D\"olz}
\ead{doelz@mathematik.tu-darmstadt.de}
\author[aff2]{H.~Harbrecht}
\ead{helmut.harbrecht@unibas.ch}
\author[aff3]{S.~Kurz}
\ead{kurz@gsc.tu-darmstadt.de}
\author[aff4]{M.~Multerer}
\ead{michael.multerer@usi.ch}
\author[aff3]{S.~Sch\"ops}
\ead{schoeps@temf.tu-darmstadt.de}
\author[aff3]{F.~Wolf{\,}\corref{corr}}
\ead{wolf@gsc.tu-darmstadt.de}

\address[aff1]{TU Darmstadt, Department of Mathematics}
\address[aff2]{Universität Basel, Department of Mathematics and Computer Science}
\address[aff3]{TU Darmstadt, Institute TEMF \& Centre for Computational Engineering}
\address[aff4]{Universit\`a della Svizzera italiana, Institute of Computational Science}
\cortext[corr]{Corresponding author}

\begin{abstract}
In this article, we present Bembel, the C++ library featuring higher order 
isogeometric Galerkin boundary element methods for Laplace, Helmholtz, and Maxwell problems.
Bembel is compatible with geometries from the Octave NURBS package, and provides an interface 
to the Eigen template library for linear algebra operations. 
For computational efficiency, it applies an embedded fast multipole method tailored to the isogeometric analysis framework
and a parallel matrix assembly based on OpenMP.
\end{abstract}


\end{frontmatter}

\section{Introduction}
The \emph{boundary element method} (BEM) or \emph{method of moments} (MoM) is a
widely used tool for the solution of partial differential
equations (PDEs) in engineering applications such as acoustic and electromagnetic scattering problems in homogeneous
media. It is accepted in industry and implementations are available as
software packages. Prominent examples are BEM++ \cite{SBA+15} which is accessible open source and BETL \cite{HK12} which is freely available for academic use.
A commonality of these codes is that they are relying on mesh generators to create
triangulation-based surface meshes consisting of flat triangles or lower order parametric
elements. Thus, the order of convergence of (higher-order) boundary element methods is limited
by the approximation error of the underlying mesh.

With the emergence of \emph{isogeometric analysis}, see \cite{HCB05}, boundary element
methods have received increased attention by the community, since \emph{computer aided 
design} (CAD) tools directly provide parametric representations of surfaces in terms 
of \emph{non-uniform rational B-splines} (NURBS). This has recently led to a
flourishing development of software concerning NURBS, see, e.g., \cite{SCFV, BK19}.
Since the extraction of volume mappings
from surface descriptions is an active research area with unsolved open problems, the use of 
isogeometric finite element methods is quite challenging in practice. Boundary element methods are the
methods of choice in this setting, see also \cite{ACD+18,DHK+18,FGHP17,MZB15,SBTR12,Tau15,TM12}. 
However, their implementation usually requires a significant amount of expert knowledge, 
which can lead non-experts to refrain from their usage.

The software library Bembel, \textbf{B}oundary \textbf{E}lement \textbf{M}ethod \textbf{B}ased 
\textbf{E}ngineering \textbf{L}ibrary, is written in C and C++ \cite{Bembel}. It solves boundary value problems 
governed by the Laplace, Helmholtz or electric wave equation within the isogeometric framework.
The development of the software started in the context of \emph{wavelet Galerkin methods} on 
parametric surfaces, see \cite{Har01}, where the integration routines for the Green's function
of the Laplacian have been developed and implemented. It was then extended to \emph{hierarchical matrices} 
($\mathcal{H}$-matrices) in \cite{HP13} and to \emph{$\mathcal{H}^2$-matrices} and higher-order B-splines in 
\cite{DHP16}. With support of B-splines and NURBS for the geometry mappings, the Laplace and Helmholtz code
became isogeometric in \cite{DHK+18}.
Finally, in \cite{DKSW18}, it has been extended to the electric field integral equation.

The publication of this software package aims at making isogeometric boundary element 
methods available for a broader audience. Therefore, we aim at an easy to use C++ API,
which streamlines the access to the underlying routines, with compatibility to the Eigen 
template library for linear algebra, see \cite{Gjo10}. 
This is achieved, while still providing black-box $\mathcal{H}^2$-compression of the
boundary element system matrices and OpenMP parallelized matrix assembly. The $\mathcal{H}^2$-compression
yields an almost linear complexities in the number of unknowns for assembly, storage requirements and
matrix-vector multiplication for the system matrix.
For the representation of geometries, Bembel features arbitrary parametric mappings, 
most prominently given as NURBS-mappings and can directly be imported from files 
generated by the Octave NURBS package \cite{SCFV}.

The structure of this document is as follows. Section \ref{sec::problems} shortly
recalls the main concepts of isogeometric boundary element methods.
Section~\ref{sec::considerations} is concerned with design considerations for the library,
whereas Section~\ref{sec::implementation} deals with the implementation. Afterwards,
in Section~\ref{sec::example}, we discuss an example program, whereas Section~\ref{sec::impact}
discusses the significance of the code. Finally, in Section~\ref{sec::concl}, we will
conclude our review.

\section{Isogeometric Boundary Element Methods}\label{sec::problems}
Bembel is able to treat potential, acoustic scattering, and electromagnetic scattering
problems governed by the well-known Laplace, Helmholtz, and electric wave equations as
stated in \ref{sec::PDE}. For the Helmholtz and the electric wave equation, Bembel
assumes non-resonant wave-numbers. 

The boundary value problems can be solved by \emph{single layer potential
ansatzes}, see \ref{sec::BIE}, which require the solution of \emph{boundary
integral equations} by means of a numerical discretization. This and similar approaches are  commonly known as  \emph{boundary element methods}. Bembel implements these through
the use of a conforming Galerkin scheme.

One of the inherent advantages of boundary element methods over classical finite element techniques in the volume is that they can
act directly on surface descriptions by NURBS from CAD programs, see \ref{sec::NURBS}. This, in
connection with corresponding spline spaces, leads to so-called \emph{isogeometric boundary
element methods}. 
Bembel implements this and assumes that these surface descriptions fulfill the requirements
stated in \ref{sec::Brep}. The spline spaces for the Galerkin method are constructed as isogeometric
multi-patch B-spline spaces, see \ref{sec::FEMspaces}. 

It is well known that in all cases solvability and uniquenes of the solution, both of the continuous problem as well as of its discrete counterpart, can be guaranteed by imposing reasonable assumptions on the boundary values, essentially guaranteeing ``physicality'' of the input data, see e.g.~\cite{BH03, Ste08}.

The computation of the system matrix requires the evaluation of singular integrals, see \cite{Ste08}.
For the numerical quadrature of these integrals, we employ regularization techniques as described in \cite{SS97}. 
The compression of the resulting densely populated system matrices is based on
the embedded fast multipole method (FMM), which is tailored to the framework of 
isogeometric analysis, see \cite{DHP16, DKSW18}, and fits into the framework of $\mathcal{H}^2$-matrices. Its particular advantage is that the matrix
compression is directly applied on the reference geometry, that is, the unit square.
Hence, the employed compression scheme profits from the inherently two-dimensional
structure of the problem. The complexity for assembly, storage requirement and
matrix-vector multiplication for the system matrix are almost linear in the number of
unknowns, see \cite{DHP16,DKSW18}.
Moreover, this compression technique provably maintains
the convergence behaviour for increasingly finer discretizations, cf.~\cite{DHP16}. 
An in-depth mathematical analysis of the implemented approach together with numerical studies based on previous versions of Bembel is available in \cite{DHK+18,DHP16,DKSW18}.

\section{Design Considerations}\label{sec::considerations}
Most modern three-dimensional boundary element codes with built-in matrix compression are
written in C or C++, which are proven to lead to efficient implementations. One of the
central aims of Bembel is to provide a computationally efficient isogeometric boundary element code
with a plain and simple user interface. Therefore, the API of Bembel is designed in
modern C++ and provides an interface for the Eigen template library for numerical linear algebra. 
This allows the user for a programming experience similar to Matlab and Octave 
and for the use of all matrix-free algorithms from the Eigen library. Particularly, all
iterative solvers provided by Eigen can be employed without further modifications.
In contrast, the low-level routines of Bembel are written in optimized C code which is
tuned towards efficiency. Shared memory parallelism by the use of OpenMP is provided and hence
aims at multicore machines.

\section{Implementation}\label{sec::implementation}

\subsection{Geometry}
With Bembel, geometry representations can
directly be imported from files generated by the Octave NURBS package \cite{SCFV}.
However, we should mention that the implementation of the \lstinline|Geometry|
class is rather flexible such that the NURBS mappings may be replaced by 
arbitrary smooth parametric mappings defined by the user.

\subsection{PDE}
The PDE to solve and the potential ansatz to use are defined by a derivation
of the base class \mbox{\lstinline|PDEproblem|.} The currently available
derivations are \lstinline|LaplaceSingle|, \lstinline|HelmholtzSingle|, and \lstinline|MaxwellSingle|,
which implement the boundary integral equations \eqref{eq::BIE} for Laplace, Helmholtz and Maxwell problems listed in \ref{sec::BIE}.
The derived classes contain all necessary information for the PDE and the boundary integral
equation under consideration, see \ref{sec::PDE} and \ref{sec::BIE}. 
Note that, for \lstinline|HelmholtzSingle| and \lstinline|MaxwellSingle|, the construction of a class instance
requires a wave number described as \lstinline|std::complex<double>|.

\subsection{Discretization}
Given a \lstinline|Geometry| object and a class inheriting from \lstinline|PDEproblem|, e.g.\ \mbox{\lstinline|LaplaceSingle|,}
an object of the template class \lstinline|Discretization<LaplaceSingle>|
describes a conforming boundary element space on refinement level \lstinline|L| and
polynomial degree \lstinline|P|. The type of the B-spline boundary element space is automatically defined
through the class inheriting from \lstinline|PDEproblem|, e.g.\ \lstinline|LaplaceSingle|.

\subsection{Boundary Values}
The load vectors of the Galerkin systems can be obtained via quadrature from a
\lstinline|std::function| object describing the boundary values.
For the Laplace case,
\lstinline|computeRhs| takes a \lstinline|Discretization| and a
\lstinline|std::function<double(Eigen::Vector3d)>)| object and returns
the load vector as an \lstinline|Eigen::VectorXd|. For the Helmholtz and Maxwell case, the return value is an \lstinline|Eigen::VectorXcd|. Note that in
this case the \lstinline|std::function| object has to be modified accordingly, i.e.\ it must accept
arguments of type \lstinline|Eigen::Vector3cd|. Bembel provides exemplary functions. They can be wrapped into this framework, e.g, via C++ lambdas.

\subsection{System Matrix}
Given a specialized \lstinline|Discretization|, e.g.\  \lstinline|Discretization<LaplaceSingle>|, 
and a polynomial degree for the $\cH^2$-compression, a corresponding specialization
of the
template class \lstinline|HierarchicalMatrix|, e.g.\  \mbox{\lstinline|HierarchicalMatrix<LaplaceSingle>|,} may
be used to assemble the compressed system matrix of the Galerkin
system. Note that the type of the PDE and the geometry are directly encoded into the discretization,
such that no further information is required here.

The template class \lstinline|HierarchicalMatrix<T>| inherits from the corresponding
specialization of the \lstinline|EigenBase| class, i.e.\ \lstinline|EigenBase<HierarchicalMatrix<T>>|. Therefore, a 
\lstinline|HierarchicalMatrix| object may use the Eigen interface and can be passed
to all functions which accept \lstinline|EigenBase<Derived>| objects as parameters.

The Eigen methods
\lstinline|rows()| and \lstinline|cols()| are specialized in order to return the dimensions.
Moreover, to facilitate the multiplication of \lstinline|Eigen::VectorX| objects,
the specialization \lstinline|Product<HierarchicalMatrix, Rhs, AliasFreeProduct>| of the 
Eigen \lstinline|Product| class is provided. In particular, the operator \lstinline|*| is overloaded
to support expressions like \lstinline|H * x|, where \lstinline|H| is a
\lstinline|HierarchicalMatrix<PDEproblem>| object, and \lstinline|x| an \lstinline|Eigen::VectorX|
object, with a fitting \lstinline|Scalar| type.

\subsection{Solution of the System of Linear Equations}
The solution of linear systems within Bembel can be performed by the
matrix-free solvers from Eigen. To that end, 
the template class \lstinline|HierarchicalMatrix|inherits the
\lstinline|Eigen::internal::traits| of an \lstinline|Eigen::SparseMatrix| with the
same \lstinline|Scalar| type as in the corresponding example from the 
Eigen documentation on matrix-free solvers. We opted for the 
\lstinline|Eigen::SparseMatrix| traits, since they seem to impose the least
requirements. Hence, in view of the implemented general matrix-vector (GEMV) product
\lstinline|H * x|, all Eigen iterative solvers which solely rely on this
operation may be used.

\subsection{Evaluation of the Solution}
Having the density vector \lstinline|rho| from the iterative solver, the solution to the
PDE in either the exterior or the interior domain can be evaluated by the function
\lstinline|Sol::evalSolution(gridpoints, rho, myDisc)|.
In here, gridpoints should be an \lstinline|Eigen::Matrix<double, Eigen::Dynamic, 3>|
matrix of \lstinline|n| three-dimensional points in the domain of interest. Depending on the PDE, \lstinline|sol| is either an \lstinline|Eigen::VectorXd|, \lstinline|Eigen::VectorXcd| or
\lstinline|Eigen::Matrix<std::complex<double>, Eigen::Dynamic, 3>| object, where each row contains the solution at the corresponding gridpoint.

\subsection{Visualization}
The \lstinline|Bembel::Vis::plotDiscretizationToVTK| functions can be used to
visualize \lstinline|rho| and the geometry in vtk-format. Their functionality is
explained extensively in the \texttt{src/examples/example\_Visualization.cpp} file.

\section{Example Program}\label{sec::example}
We discuss a full example for the solution of an electromagnetic scattering
problem, which is located in \texttt{src/examples/example\_MaxwellSingle.cpp}.

\begin{lstlisting}
int main() {
  using namespace Bembel;
  using namespace Eigen;

  const int multipoleDegree = 16;
  const int knotRepetition = 1;
  const std::complex<double> wavenumber(3, 0);

  Geometry geom("geometry.dat");
  MaxwellSingle pde(wavenumber);
  MatrixXd gridpoints = Util::makeSphereGrid(6, 10);

  const std::function<Vector3cd(Vector3d, std::complex<double>)>
      fun = ... ; // Define incident wave

  for (int P = 1 ; P < 10 ; ++P)) {
    for (int L = 0 ; L < 6 ; ++L) {
      Discretization<MaxwellSingle> disc(geom, pde, P, knotRepetition, L);
      VectorXcd rhs = Rhs::computeRhs(disc, fun);
      HierarchicalMatrix<MaxwellSingle> hmat(disc, multipoleDegree);
      GMRES<HierarchicalMatrix<MaxwellSingle>, IdentityPreconditioner>  gmres;
      gmres.compute(hmat);
      VectorXcd rho = gmres.solve(rhs);
      MatrixXcd pot = Sol::evalSolution(gridpoints, rho, disc);
    }
  }
  return (0);
}
\end{lstlisting}

At the beginning of the program, 
the polynomial degree for the $\cH^2$-com\-pres\-sion, the knot repetition
of the B-spline spaces, and the wave number are defined. The \lstinline|Geometry| object is
loaded from the file \lstinline|"geometry.dat"|, and the PDE is defined for a given wave number.
\lstinline|Util::makeSphereGrid(6,10)| generates ten evaluation points on ten lines of latitude for the scattered
field on a sphere around the origin with radius 6. Afterwards, an incident wave for the
scattering problem is defined in \lstinline|fun|.

The rest of the example program loops over different polynomial degrees \lstinline|P| and
refinement levels \lstinline|L|. In the loop, the discrete space is defined as a
\lstinline|Discretization| object, the load vector and the system matrix are assembled and solved with the
GMRES solver from the Eigen (unsupported) package. Finally, the scattered field is evaluated
at the previously defined points.

\begin{figure}
	\input{figures/include.tex}
	\centering
	\begin{minipage}{0.46\textwidth}
		\includegraphics[width=\textwidth]{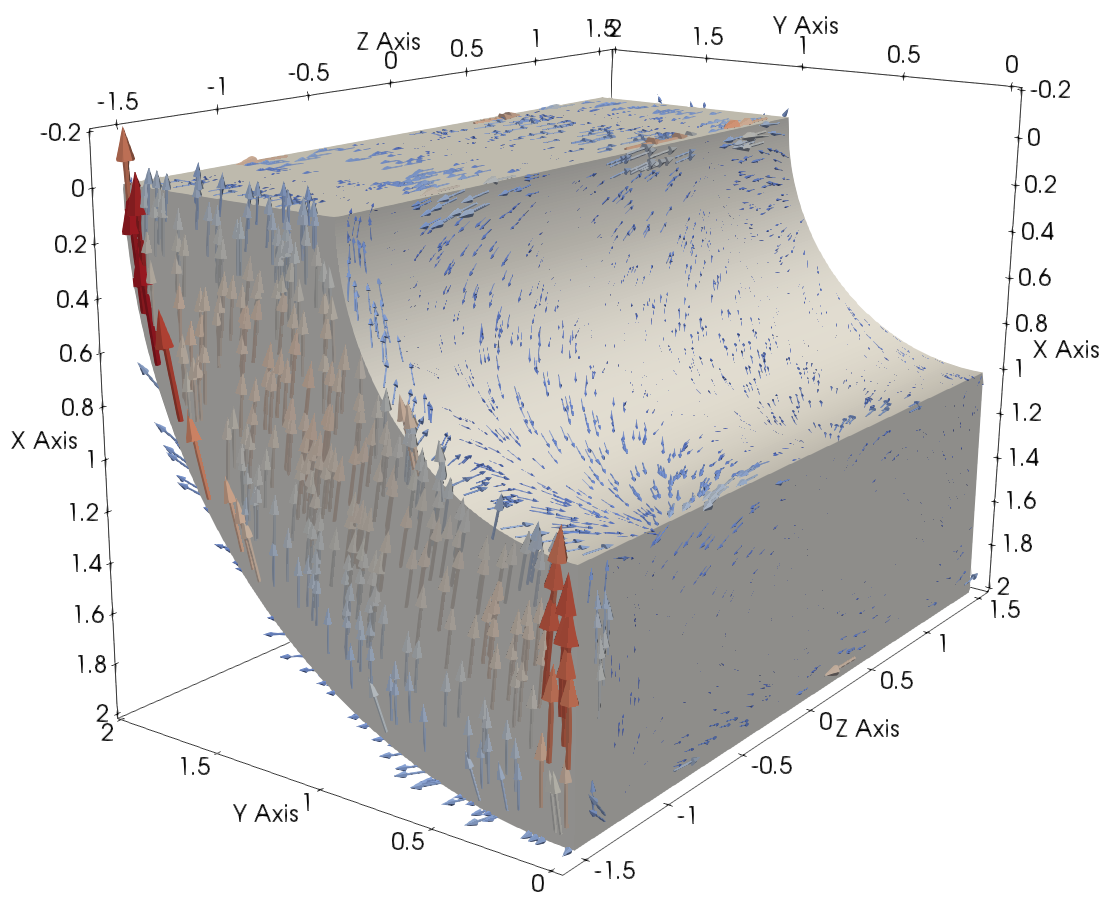}
	\end{minipage}
	\begin{minipage}{0.5\textwidth}
	\scalebox{0.55}{
	\begin{convplot}
		\convline{figures/test/MaxwellSingle_1.log}{\lstinline|P| = 1}
		\convline{figures/test/MaxwellSingle_2.log}{\lstinline|P| = 2}
		\convline{figures/test/MaxwellSingle_3.log}{\lstinline|P| = 3}
		\convfitorder{figures/test/MaxwellSingle_1.log}{3}
		\convfitorder{figures/test/MaxwellSingle_2.log}{5}
		\convfitorder{figures/test/MaxwellSingle_3.log}{7}
	\end{convplot}
	}
	\end{minipage}
	\caption{Illustrations obtained from an electromagnetic problem. Quarter-pipe geometry with
		illustrated surface current (i.e.,~\lstinline|rho|) of a plane
		electromagnetic wave in longitudinal direction (left). Convergence plots generated by the provided \LaTeX-files
		illustrate how the rate of convergence 	increases with the degree \lstinline|P| of the isogeometric basis
		functions (right). The convergence plots were computed around the
		pictured geometry, with a dipole in direction $(0,0,1/100)^{\intercal}$ placed at $(1,1,0)^\intercal$, see \cite{DKSW18}
		and the example programs. The rest of the parameters are chosen as in the listed program.
		\label{fig::convplot}}
\end{figure}

Note that the program is straightforwardly adapted to the Laplace and Helmholtz cases as indicated
by the example files in the \texttt{src/examples} folder. These examples also provide error measurements of
the solution to the underlying PDE and output of these errors into log-files. These were not included to the above
example for a better understanding of the main functionality of the code.
\LaTeX-files creating error plots such as illustrated in Figure~\ref{fig::convplot} can be found in the
\texttt{LaTeX} folder. The routines can visualize the intermediate quantity \texttt{rho}
as illustrated in Figure~\ref{fig::convplot}.

\section{Impact}\label{sec::impact}

The implementation of boundary element methods in three spatial dimensions is a non-trivial
task due to the necessary numerical evaluation of singular integrals and the
required matrix compression to achieve computational efficiency. While this is already
true for lowest-order implementations for the Laplace equation on boundary triangulations with flat elements, 
difficulties increase on isogeometric (or parametric) surfaces, 
isogeometric B-spline (or higher-order) boundary element spaces, and more involved electromagnetic problems. These
implementation-related issues lead many people to refrain from the use of boundary
element methods, even when a specific engineering problem is known to be solved best
therewith.
Our software package aims to provide a state-of-the-art boundary element toolbox for engineers who would
like to apply competitive isogeometric boundary element methods in a black-box fashion.
This allows users to freely employ boundary element methods as a tool in involved engineering applications. 
To the best of our knowledge, the functionality of higher-order B-spline boundary element spaces is the first 
open source implementation available for three dimensions.

A major strongpoint of Bembel is the direct integration into the Eigen Linear Algebra
Library. This allows the user for a straightforward pre or post processing of data
with a clean user interface. The matrix-free algorithms of Eigen provide different
kinds of iterative solvers for the Galerkin systems or eigenvalue problems.

\section{Conclusion}\label{sec::concl}

Bembel is an open-source library enabling users to apply isogometric boundary element methods for potential, acoustic, electromagnetic, and many other problems in a black-box fashion.
This is achieved through an easy-to-use API and compatibility with the Eigen 
template library for linear algebra.
Moreover, it encorporates state-of-the-art compression techniques for large problems, 
as well as OpenMP parallelization.

\section*{Acknowledgement}
\begin{footnotesize}
This work is supported by DFG Grants SCHO1562/3-1 and KU1553/4-1 within the project \emph{Simulation of superconducting cavities with isogeometric boundary elements (IGA-BEM).}
The work of J\"urgen D\"olz was partially supported by SNSF Grants 156101 and 174987, as well as the Graduate School of Computational Engineering at TU Darmstadt and the Excellence Initiative of the German Federal and State Governments and the Graduate School of Computational Engineering at TU Darmstadt.
Michael Multerer was partially supported by SNSF Grant 137669.
The work of Felix Wolf is supported by the Excellence Initiative of the German Federal and State Governments and the Graduate School of Computational Engineering at TU Darmstadt.
\end{footnotesize}

\begin{table}[!h]
\section*{Metadata}
\begin{tabular}{|l|p{5cm}|p{5cm}|}
\hline
\textbf{Nr.} & \textbf{Code metadata description} & \textbf{Please fill in this column} \\
\hline
C1 & Current code version & v0.9 \\
\hline
C2 & Permanent link to code/repository used for this code version & \href{http://www.bembel.eu/}{http://www.bembel.eu}, cf. \cite{Bembel}\\
\hline
C3 & Legal Code License   & GPL3 \\
\hline
C4 & Code versioning system used & Git \\
\hline
C5 & Software code languages, tools, and services used & C, C++, OpenMP \\
\hline
C6 & Compilation requirements, operating environments \& dependencies & CMake, C++11, Eigen Linear Algebra Library\\
\hline
C7 & If available Link to developer documentation/manual & \href{http://temf.github.io/bembel/assets/DOC.html}{http://www.bembel.eu/ \mbox{$\rightarrow$ Documentation}}\\
\hline
C8 & Support email for questions & {info@bembel.eu} \\
\hline
\end{tabular}
\caption{Code metadata}
\label{} 
\end{table}

\section*{References}
\bibliography{mybibfile}
\label{}

\appendix

\section{Partial Differential Equations}\label{sec::PDE}
In the following, let $\Omega\subset\mathbb{R}^3$ be a bounded domain with Lipschitz boundary $\Gamma\isdef\partial\Omega$.
Moreover, we define the exterior domain $\Omega^{\operatorname{c}}\isdef\R^3\setminus\overline{\Omega}$.
Bembel can treat the Laplace equation
\begin{align}
\begin{aligned}
-\Delta u_\mathrm{L} = {}&{}0&&\text{in}~\Omega,\\
u_\mathrm{L} = {}&{}g_\mathrm{L}&&\text{on}~\Gamma,
\end{aligned}\tag{LP}
\label{eq::laplace}	
\end{align}
and the Helmholtz equation
\begin{align}
\begin{aligned}
-\Delta u_\mathrm{H} - \kappa_\mathrm{H}^2u_\mathrm{H}= {}&0&&\text{in }\Omega^{\operatorname{c}},\\
u_\mathrm{H} = {}&g_\mathrm{H}&&\text{on }\Gamma,
\end{aligned}\tag{HP}\label{eq::helmholtz}
\end{align}
with Sommerfeld radiation conditions towards infinity. 
In both cases, the boundary values have to be understood in the usual sense of traces, 
see, e.g. \cite{Ste08}. In addition, Bembel can treat the electric wave equation
\begin{align}
\begin{aligned}
\bcurl\bcurl\bfE_\mathrm{M}-\kappa_\mathrm{M}^2\bfE_\mathrm{M}={}&\bfNull&&\text{in}~\Omega^{\operatorname{c}},\\
\bfn\times\bfE_\mathrm{M}={}&\bfg_\mathrm{M}&&\text{on }\Gamma,
\end{aligned}\tag{MP}\label{eq::maxwell}
\end{align}
with Silver-M\"uller radiation conditions towards infinity, which can be derived from Maxwell's equations. 
Again, the boundary values
have to be understood in the sense of traces, see \cite{BH03}.

\section{Boundary Integral Equations}\label{sec::BIE}
The three boundary value problems \eqref{eq::laplace}, \eqref{eq::helmholtz}, and \eqref{eq::maxwell} can each be solved 
by means of a \emph{single layer potential ansatz}, i.e., setting
\begin{align*}
u_\mathrm{L}(\bfx)={}&(\tilde{\cS}_\mathrm{L}\rho_\mathrm{L})(\bfx)\isdef\int_{\Gamma}\frac{1}{4\pi\|\bfx-\bfy\|_2}\rho_\mathrm{L}(\bfy)\d\sigma_{\bfy},\qquad\bfx\in\Omega,
\tag{LS}\label{eq::laplacesingle}\\
u_\mathrm{H}(\bfx)={}&(\tilde{\cS}_\mathrm{H}\rho_\mathrm{H})(\bfx)\isdef\int_{\Gamma}\frac{e^{-i\kappa_\mathrm{H}\|\bfx-\bfy\|_2}}{4\pi\|\bfx-\bfy\|_2}\rho_\mathrm{H}(\bfy)\d\sigma_{\bfy},\qquad\bfx\in\Omega^c,
\tag{HS}\label{eq::helmholtzsingle}\\
\bfE_\mathrm{M}(\bfx)={}&(\tilde{\bfcS}_\mathrm{M}\bfj_\mathrm{M})(\bfx)
\isdef{}\int_{\Gamma}\frac{e^{-i\kappa_\mathrm{M}\|\bfx-\bfy\|_2}}{4\pi\|\bfx-\bfy\|_2}\bfj_\mathrm{M}(\bfy)\d\sigma_{\bfy}&&\\
&\qquad\qquad\qquad+\frac{1}{\kappa_\mathrm{M}^2}\bgrad_{\bfx}\int_{\Gamma}\frac{e^{-i\kappa_\mathrm{M}\|\bfx-\bfy\|_2}}{4\pi\|\bfx-\bfy\|_2}\div_{\Gamma}\bfj_\mathrm{M}(\bfy)\d\sigma_{\bfy},\quad\bfx\in\Omega^c.
\tag{MS}\label{eq::maxwellsingle}
\end{align*}
It can be shown that \eqref{eq::laplacesingle}, \eqref{eq::helmholtzsingle}, and
\eqref{eq::maxwellsingle} solve the boundary value problems \eqref{eq::laplace},
\eqref{eq::helmholtz}, and \eqref{eq::maxwell} for appropriate density functions
$\rho_\mathrm{L}$, $\rho_\mathrm{H}$, and $\bfj_\mathrm{M}$, see \cite{BH03,Ste08}.
Taking the proper trace operators one arrives at the three boundary integral equations
\begin{align}\label{eq::BIE}
\cS_\mathrm{L}\rho_\mathrm{L}=g_\mathrm{L}, \quad
\cS_\mathrm{H}\rho_\mathrm{H}=g_\mathrm{H},\quad
\bfcS_\mathrm{M}\rho_\mathrm{M}=\bfg_\mathrm{M}\tag{BIE}
\end{align}
on $\Gamma$ to determine the unknown density functions in 
\eqref{eq::laplacesingle}, \eqref{eq::helmholtzsingle}, and \eqref{eq::maxwellsingle}. We note that $\cS_\mathrm{L}$ and
$\cS_\mathrm{H}$ are isomorphisms from $H^{-1/2}(\Gamma)$ to $H^{1/2}(\Gamma)$, whereas $\bfcS_\mathrm{M}$ is an isomorphism
on $\bfH_{\times}^{-1/2}(\div_{\Gamma},\Gamma)$, see \cite{BH03,Ste08} for
details. Thus, in view of a conforming Galerkin method, piecewise polynomial boundary element spaces are sufficient for the boundary element based solution of \eqref{eq::laplace} and \eqref{eq::helmholtz}, whereas
\eqref{eq::maxwell} requires divergence conforming boundary element spaces.
Unique solvability of the corresponding linear systems can be proven, see \cite{BH03,Ste08}.

\begin{remark}
	In fact, the described single layer potential approaches solve the interior and the
	exterior problems of \eqref{eq::laplace}, \eqref{eq::helmholtz}, and \eqref{eq::maxwell} simultaneously. Thus, $\tilde{\cS}_\mathrm{H}\rho_\mathrm{H}$ and $\tilde{\bfcS}_\mathrm{M}\bfj_\mathrm{M}$ satisfy \eqref{eq::helmholtz} and \eqref{eq::maxwell} also in $\Omega$, whereas $\tilde{\cS}_\mathrm{L}\rho_\mathrm{L}$ satisfies \eqref{eq::laplace} also in $\Omega^c$, with additional radiation conditions
	$|u_\mathrm{L}(\bfx)|=\mathcal{O}\big(\|\bfx\|_2^{-1}\big)$ and
	$\|\bgrad u_\mathrm{L}(\bfx)\|_2=\mathcal{O}\big(\|\bfx\|_2^{-2}\big)$ towards infinity, see \cite{BH03,Ste08}.
\end{remark}

\section{B-Splines and NURBS}\label{sec::NURBS}
Let $p$ and $k$ be two fixed integers such that $0\leq p< k$ and let $\Xi$ be a locally quasi uniform knot
vector with knots in $[0,1]$, see \cite{Piegl_1997aa}.
The B-spline basis $ \lbrace b_j^p \rbrace_{0\leq j< k}$ is then defined by recursion as
\begin{align*}
b_j^p(x) & =\begin{cases}
\chi_{[\xi_j,\xi_{j+1})},&\text{ if }p=0,\\[8pt]
\frac{x-\xi_j}{\xi_{j+p}-\xi_j}b_j^{p-1}(x) +\frac{\xi_{j+p+1}-x}{\xi_{j+p+1}-\xi_{j+1}}b_{j+1}^{p-1}(x), & \text{ else,}
\end{cases}
\end{align*}
where $\chi_M$ denotes the indicator function for the set $M$. Having the B-spline basis at our disposal, 
we define the spline space  $S^p(\Xi)\isdef\operatorname{span}(\lbrace b_j^p\rbrace_{0\leq j <k})$.

A NURBS mapping $\boldsymbol\gamma_i\colon\square\to\mathbb{R}^3$ on the unit square $\square=[0,1]^2$ is given by
\begin{align*}
\boldsymbol\gamma_j(x,y)\isdef \sum_{0\leq j_1<k_1}\sum_{0\leq j_2<k_2}\frac{\bfc_{j_1,j_2} b_{j_1}^{p_1}(x) b_{j_2}^{p_2}(y) w_{j_1,j_2}}{ \sum_{i_1=0}^{k_1-1}\sum_{i_2=0}^{k_2-1} b_{i_1}^{p_1}(x) b_{i_2}^{p_2}(y) w_{i_1,i_2}},
\end{align*}
and described by its control points $\bfc_{j_1,j_2}\in \R^3$ and weights $w_{i_1,i_2}>0.$ For further concepts and algorithmic realization of the NURBS, we refer to \cite{Piegl_1997aa}.

\section{Boundary Representation}\label{sec::Brep}
Bembel assumes that the boundary representations are the union of several \emph{patches} $\Gamma_i$, i.e.,
\begin{align*}
\boldsymbol\gamma_i:\square\to\Gamma_i\quad\text{ with }\quad\Gamma_i = \boldsymbol\gamma_i(\square)
\quad\text{ for } i = 1,2,\ldots,M,
\end{align*}
where $\boldsymbol\gamma_i$ is given by a NURBS mapping with outward pointing normal. The boundary itself is then
the collection of all patches
\[
\Gamma = \bigcup_{i=1}^M \Gamma_i,
\]
where the intersection $\Gamma_i\cap\Gamma_{i^\prime}$ 
consists at most of a common vertex or a common edge for 
\(i\neq i^\prime\).

In order to ensure conforming meshes,
Bembel also imposes the following matching condition on the parametrizations:
For each \({\bf x}=\boldsymbol\gamma_{i}({\bf s})\)
on a common edge of \(\Gamma_i\) and \(\Gamma_{i'}\), there has to exist
a bijective and affine mapping \(\boldsymbol{\Xi}\colon\square\to\square\) 
such that there holds
\(\boldsymbol\gamma_{i}({\bf s})=(\boldsymbol\gamma_{i'}\circ \boldsymbol{\Xi})({\bf s})\).
This means that the parameterizations \(\boldsymbol\gamma_i\) and \(\boldsymbol\gamma_{i'}\) 
coincide on the common edge except for orientation.

\section{B-Spline Boundary Element Spaces}\label{sec::FEMspaces}
The boundary element spaces for the Galerkin method are constructed as isogeometric multi-patch B-spline spaces, see \cite{BDK+18}.
Bembel uses equidistant knot vectors $\Xi_1=\Xi_2$ with $2^L$ elements and the same polynomial degrees in each direction, 
i.e., $p_1=p_2$, on the unit square. The corresponding spline
spaces are thus given by
\begin{align*}
\S^0_{\bfp,\boldsymbol\Xi}(\square) &{}= S_{p_1,\Xi_1}([0,1]) \otimes S_{p_2,\Xi_2}([0,1]),\\
\boldsymbol{\mathbb{S}}^1_{\bfp,\boldsymbol\Xi}(\square) &{}= S_{p_1,\Xi_1}([0,1]) \otimes S_{p_2-1,\Xi_2'}([0,1]) \times S_{p_1-1,\Xi_1'}([0,1]) \otimes S_{p_2,\Xi_2}([0,1]),\\
\S^2_{\bfp,\boldsymbol\Xi}(\square) &{} = S_{p_1-1,\Xi_1'}([0,1]) \otimes S_{p_2-1,\Xi_2'}([0,1]),
\end{align*}
for $\bfp=(p_1,p_2)$ and $\boldsymbol\Xi = (\Xi_1,\Xi_2)$. Herein, $\Xi_i'$, for $i=1,2$, denotes the truncated knot vector,
i.e., the knot vector without its first and last element.

Due to the representation of the computational geometry by patches, the spaces constructed on the unit square
can easily be lifted to the patches on the boundary by the \emph{Piola transform}. 
Enforcing continuity conditions across patch boundaries yields conforming discretizations
\begin{align*}
\S^0_{\bfp,\boldsymbol\Xi}(\Gamma)\subset H^{1/2}(\Gamma),\quad
\boldsymbol{\S}^1_{\bfp,\boldsymbol\Xi}(\Gamma)\subset \bfH_{\times}^{-1/2}(\div_{\Gamma},\Gamma),\quad
\S^2_{\bfp,\boldsymbol\Xi}(\Gamma)\subset H^{-1/2}(\Gamma).
\end{align*}

\end{document}